\documentstyle[preprint,aps]{revtex}

\begin{document}
\draft
\preprint{LBL-37388/SNUTP-95-070}
\title{Thermal effects on dilepton production from $\pi-\pi$ annihilation}
\bigskip
\author{Chungsik Song$^1$, Volker Koch$^2$, Su Houng Lee$^3$ and C. M.
Ko$^1$}
\address{\it
$^1$Cyclotron Institute and Physics Department\\
 Texas A\&M University, College Station, TX 77843, USA \\
$^2$Nuclear Science Division, Lawrence Berkeley Laboratory,\\
 University of California, Berkeley, CA 94720, USA\\
$^3$Department of Physics, Yonsei University, Seoul 120-749, Korea}
\date{\today}
\maketitle
\begin{abstract}

We study finite temperature effects on dilepton production from
pion-pion annihilation in hot hadronic matter.
The softening of the pion dispersion relation in a medium is found to
enhance the production rate of dileptons with invariant masses in the
region of $2m^*_\pi(T)<M<m_\rho$.  On the other hand, the reduction of the
pion electromagnetic form factor at finite temperature leads to
a suppression of the dilepton production rate.
Including both effects, we have found that the dilepton yield
is slightly enhanced
in the invariant mass region $M= 270\sim 600$ MeV
but is suppressed around the vector meson resonance.
We further discuss the relevance of our results to
recent experimental data from the S+Au collisions at
CERN/SPS energies by the CERES collaboration.
\end{abstract}
\eject

\newpage

Dilepton production from relativistic heavy ion collisions
has continually attracted great interest for possible
signals of the quark-gluon plasma that is expected to be formed in
the initial stage of the collision \cite{dilepton,shuryak,ruu,kox}.
Since dileptons would escape from the collision region without further
interactions, they carry the information about the hot dense matter
from which they are produced and, therefore, are considered as
reliable probes for the hadronic mater at extreme condition.
It has been suggested that a window for observing dileptons from the
plasma phase exists in the invariant mass region
between the $\phi$ and $J/\psi$ \cite{shuryak}.
Above the $J/\psi$ mass, dominant
contributions are from the Drell-Yan process and direct charm decay \cite{ruu},
while below the $\phi$ meson mass, radiative and
direct decays, together with $\pi\pi$ annihilation, form the most important
sources \cite{hadron}.

Recently, experimental data on dileptons have been measured by
the CERES collaboration at the CERN/SPS \cite{dilep-e}. In proton induced
reactions, such as the 450 GeV $p-Be$ and $p-Au$ collisions, the low-mass
dilepton spectra can be satisfactorily explained by dileptons
from hadron decays. On the other
hand, in the $S+Au$ collision at 200 GeV/nucleon a significant enhancement
of dileptons over the hadronic contribution
has been observed in the invariant mass region 200 MeV $<M<$ 1 GeV.
The enhancement seen in the CERES experiment is for dileptons at the central
rapidity where the charge particle density is high. In another experiment
by the HELIOS-3 collaboration \cite{helios}, dileptons at forward rapidities,
where the charge particle density is low, were measured, and the enhancement
was found to be smaller. Suggestions have thus been made that the excess
dileptons seen in these experiments are from
pion-pion annihilation, $\pi^+\pi^-\to e^+e^-$.
However, the inclusion of this contribution in model calculations cannot
explain the observed dilepton yield as long as the production rate is
calculated with parameters in free space \cite{gqli,denish}.

In this letter, we study the modification of pion properties at finite
temperature and its effect on dilepton production from pion-pion
annihilation in hot hadronic matter.
To study the pion properties at finite temperature
we shall use an effective chiral Lagrangian
that includes explicitly vector mesons.
In the literature, two methods have been introduced to include vector
mesons and photon field in the chiral Lagrangian;
the massive Yang-Mills approach \cite{MYMA}
and the hidden gauge approach\cite{BKY}.
These two methods have been shown to be gauge equivalent and to
have identical symmetry properties at finite temperature \cite{slee}.

We shall follow the hidden gauge approach by considering the
$[SU(2)_{\rm L} \times SU(2)_{\rm R}]_{\rm global}$ $\times$
$[SU(2)_V]_{\rm local}$ ``linear" model.  It is constructed with two
SU(2)-matrix valued variables $\xi_L(x)$ and $\xi_R(x)$, which transform
as $\xi_{L,R}(x) \rightarrow \xi'_{L,R}(x)=h(x) \xi_{L,R}~ g^\dagger_{L,R}$
under
$h(x)\in[$SU(2)$_V]_{\rm local}$ and $g_{L,R}\in[$SU(2)$_{L,R}]_{\rm global}$.
Introducing the vector meson $V_\mu$  as the gauge field of the local
symmetry and the photon ${\cal B}_\mu$ as an external gauge field of the global
symmetry, we have the following chirally invariant Lagrangian,
\begin{eqnarray}
{\cal L}& = & f_\pi^2 {\rm tr} \left[\frac{1}{2i} ( {\cal D}_\mu \xi_L \cdot
         \xi_L^\dagger - {\cal D}_\mu \xi_R \cdot \xi_R^\dagger) \right]^2
 \nonumber \\ [12pt]
      & & + af_\pi^2 {\rm tr} \left[gV_\mu-\frac{1}{2i} ( {\cal D}_\mu \xi_L
      \cdot \xi_L^\dagger + {\cal D}_\mu \xi_R \cdot \xi_R^\dagger)  \right]^2
      + {\cal L}_{\rm kin} ( V_\mu ,{\cal B}_\mu),
\end{eqnarray}
where $f_\pi=93$ MeV is the pion decay constant.
The covariant derivative, ${\cal D}_\mu \xi_{L,R}$, is given by
\begin{eqnarray}
{\cal D}_\mu \xi_{L,R}= \partial_\mu \xi_{L,R}
+ ie \xi_{L,R} {\cal B}_\mu \tau_3/2.
\end{eqnarray}
We also add a term that explicitly breaks the chiral symmetry,
\begin{equation}
{\cal L}_{SB}={1\over4}f_\pi^2 m_\pi^2
{\rm tr} (\xi_L\xi_R^\dagger+\xi_R\xi_L^\dagger).
\end{equation}
In the ``unitary" gauge
\begin{equation}
\xi^\dagger_L(x)=\xi_R(x)\equiv\xi(x)=\exp(i\pi(x)/f_\pi),
\end{equation}
and with $a=2$, this effective Lagrangian
is known to give the universality of $\rho$-couplings ($g_{\rho\pi\pi}=g$),
the KSRF relations ($m_\rho^2=2f_\pi^2g^2_{\rho\pi\pi}$), and
the $\rho$ meson dominance of the
pion electromagnetic form factor ($g_{\gamma \pi \pi}=0$).
The Lagrangian has been extended to include the anomalous interactions
and axial vector mesons. However, these effects are not included in
present calculations.


First we consider the modification of the pion dispersion relation
in hot hadronic matter which is in thermal equilibrium at temperature $T$.
In effective Lagrangian approaches
at finite temperature, we assume that known hadronic
interactions can be extrapolated to finite temperature and describe
the interactions among particles in hot hadronic matter.
The temperature effect on the properties of pions can then be
studied via thermal loop-corrections which are due to interactions with
particles in the hot matter. We include only one-loop diagrams as the
hadronic matter is rather dilute at temperatures considered here.

The dispersion relation is found by locating the poles of
the propagator and is determined by an equation of the form
\begin{equation}
\omega^2={\vec k}^2+m_\pi^2+\Pi(\omega,{\vec k}),
\label{dispersion}
\end{equation}
where $\omega$ and ${\vec k}$ are the pion energy and momentum, respectively.
The self-energy $\Pi$ includes the modification of pion properties due to
interactions with particles in the hot matter. While the real part gives
its dispersion relation, the imaginary part of
the self-energy determines the absorption of pions in hot matter.

The self-energy of a pion at finite temperature is calculated from
diagrams in fig.~1 at the one-loop order.
The contributions from fig.~1.a ($\Pi^{(a)}$) and 1.b ($\Pi^{(b)}$)
are given, respectively, by
\begin{eqnarray}
\Pi^{(a)}(\omega, \vec k) &=& 2g^2T\sum_n\int{d^3p\over(2\pi)^3}
{(p_\mu-k_\mu)(g^{\mu\nu}-(p^\mu+k^\mu)(p^\nu+k^\nu)/m_\rho^2)
(p_\nu-k_\nu)\over (p^2-m_\pi^2)((p+k)^2-m_\rho^2)},
\nonumber \\ [12pt]
\Pi^{(b)}(\omega, \vec k)&=& T\sum_n\int{d^3p\over(2\pi)^3}
\left({5\over6}{m_\pi^2\over f_\pi^2}+{1\over 3f_\pi^2}(p^2+k^2)\right)
{1\over (p^2-m_\pi^2)},
\end{eqnarray}
where $k$ and $p$ are the external and loop pion four momenta, respectively,
and $p^2=p_0^2-{\vec p}^2$ with $p_0=2\pi nTi$.
Eq.~(\ref{dispersion}) is now solved self-consistently with
\begin{equation}
\Pi(\omega,\vec k)=\Pi^{(a)}(\omega,\vec k)+\Pi^{(b)}(\omega,\vec k).
\end{equation}
We find that the pion dispersion relation is
slightly modified at finite temperature, and the effective mass of
a pion,  which is defined as the pole position of the pion propagator in
medium,  decreases with the temperature, $m_\pi^*\approx 133$ MeV at
$T=160$ MeV. A similar result has been obtained in massive Yang-Mills approach
\cite{song_pion}.


Using the temperature-dependent pion dispersion relation and effective mass,
we have calculated the dilepton production rate, $R={dN^{e^+e^-}/dx^4}$,
from pion-pion annihilation in hot hadronic matter.
For simplicity, we carry out the calculation in the rest
frame of the virtual photon.
The production rate of dileptons with vanishing three momentum
in hot hadronic matter is then given by \cite{gale}
\begin{equation}
{d^4R\over dMdq^3}\Biggr\vert_{\vec{q}=0}=
{\alpha^2\over3(2\pi)^4}
{\vert F_\pi(M)\vert^2\over (e^{M/2T}-1)^2}
\sum_{\rm k}{{\rm k}^4\over\omega^4}
\Biggl\vert{d\omega\over d{\rm k}}\Biggr\vert^{-1}
\label{dilep}
\end{equation}
where ${\rm k}=\vert {\vec k}\vert$ and $M$ is the dilepton invariant mass.
The momentum ${\rm k}$ and energy $\omega$ of the pion are related by its
dispersion relation in the medium, and the last factor in the above equation
takes into account this effect.
The sum over ${\rm k}$'s is restricted by $\omega({\rm k})=M/2$.
The pion electromagnetic form factor is denoted by $F_\pi(M)$
and is given by
\begin{equation}
\vert F_\pi(M)\vert^2={m_\rho^4\over (M^2-m_\rho^2)^2+m_\rho^2\Gamma_\rho^2},
\end{equation}
with the rho meson width $\Gamma_\rho=152$ MeV.
\footnote{In principle the formfactor is given by the more sophisticated
 Gounaris-Sakurai formula \cite{gounaris}. However, since we are only
 interested in relative changes due to in medium corrections, the simpler from
 used here  is sufficient.}

The dilepton production rate from pion-pion annihilation is shown in fig.~2
for $T=$ 160 MeV.
The result obtained with the modified pion dispersion relation
(dashed line) is
compared with that calculated using the dispersion relation
in free space (dotted line).
There are two prominent effects due to changes of the
pion properties in hot hadronic matter.
First, the threshold of dilepton production
from pion-pion annihilation is lowered because of the reduction of the
pion mass at finite temperature.
Secondly, the dilepton production rate is
enhanced in the invariant mass region, $2m^*_\pi(T)<M<m_\rho$,
and shows a peak at $M\sim 350$ MeV,
which is due to the softening of the pion dispersion relation in medium.
The latter, however, does not have any effect on dileptons with invariant
masses
near the vector meson resonance.

To ensure gauge invariance, we need to include the
temperature dependence of the pion electromagnetic from factor
as pointed out in Refs. \cite{pratt}
for the case of finite density.
The temperature dependence of the form factor involves
the modification of the rho-pion-pion vertex,
the change in rho meson properties, and the correction to
the rho-photon coupling in hot matter.  These effects have been
studied in Ref. \cite{song2} with the same Lagrangian, and it has been shown
that the production rate of dileptons with invariant masses
around the vector resonance
is suppressed.  We have repeated the calculation of
Ref. \cite{song2} by including
the finite pion mass which was neglected in the previous
calculation.
The result with the temperature-dependent pion electromagnetic form factor
is shown in fig.~2 by the dash-dotted line.  The modification of the pion
electromagnetic form factor at finite temperature is seen to reduce
the dilepton production rate.
The result obtained with both the
form factor and the dispersion relation evaluated at finite temperature
is given by the solid line.
Compared to the free case (dotted line) the combined medium effects yield
a very small enhancement close to the two pion threshold and a strong
reduction in the rho meson mass region. The ratio between the yield at
the two pion threshold and that at the rho meson mass, however, is
enhanced by a factor of 4 due to medium modifications of the pion properties.

The above calculations have been done for dileptons with vanishing three
momentum.
To compare with experimental data, we need to extend
the calculations to dileptons with
finite three momentum.  For free pions the
production rate for dileptons with a momentum $\vec q$ is given by
\begin{equation}
{d^4R\over dMdq^3}\Biggr\vert_{\vec{q}\ne 0}=
{exp[-(\sqrt{q^2+M^2}-M)/T]\over \sqrt{1+q^2/M^2}}
\left({dR\over dMdq^3}\Biggr\vert_{\vec{q}= 0}\right).
\label{result}
\end{equation}
In medium the result is nontrivial, and
we have thus assumed that the above equation remains valid.
This seems to be reasonable as
dominant contributions to the pion electromagnetic form factor
at finite temperature are momentum-independent
and also the pion in-medium dispersion relation is not very different
from the free one.
By factorizing the momentum dependence in this way one can easily
integrate over the dilepton three-momentum,
and the result is shown in fig.~3.
Because of the finite momentum effect, the dilepton spectrum obtains a quite
different shape;
the production rate in the low invariant mass region is reduced,
and the peak near $M\sim 2m_\pi$ disappears.
Even though the production rate is enhanced for dileptons in the
invariant mass region near the two pion threshold, the absolute
yield is now less than that of the vector resonance peak.


In summary, we have studied the pion properties
in hot hadronic matter
and the dilepton production rate from pion-pion annihilation with
the modified pion properties. It is shown that the
production rate is suppressed in the invariant mass region
near the vector meson resonance but slightly
enhanced at $M\sim 2m_\pi$.
The suppression mainly comes from the reduction of the pion
electromagnetic form factor but is recovered near the two pion threshold
due to modifications of the pion dispersion relation in hot matter.
After integrating over the space-momentum of the virtual photon
we obtain a quite different shape of the dilepton spectrum
compared with that of zero space-momentum dileptons.
Here, we have assumed the same momentum dependence as that for free pions but
this approximation might be too simple to take into account
the full effects due to finite dilepton momenta.
More work is thus needed to include more rigorously the finite momentum
effect.

The dilepton production rate from the $\pi-\pi$ contribution
has been calculated using a relativistic transport model
and compared with the data from S+Au collisions at the CERN/SPS \cite{gqli}.
The model calculation shows, however, a disagreement with the experimental
data unless effects due to the reduced vector meson masses in medium
are included. The estimated dilepton yield
with free vector meson masses
is about a factor $4\sim 6$ less than
the measured one in the low invariant mass region of $M=$ 200 $\sim$
600 MeV but about a factor 2 more near the vector meson resonance.
A similar result has also been obtained in hydrodynamic model
calculations including the quark-hadron phase transition \cite{denish}.
In this calculation, the most important contribution to dilepton
production comes from the hadronic component of the mixed phase at $T_c=160$
MeV.

The thermal effects on dilepton production from the early stage of
the slowly expanding
hot dense hadronic matter or from the hadronic
component of the mixed phase can
be easily estimated from eq.~(\ref{result}) and fig.~3.
Our results indicate that the production rate of dileptons
near the vector meson mass will be suppressed by the reduction of the pion
electromagnetic form factor at finite temperature.
The thermal effect on the pion dispersion relation also shows an enhancement
of dilepton productions with invariant masses $M=$ 270 MeV $\sim$ 600 MeV
and brings the shape of the dilepton spectrum close to the one measured.
However, thermal effects may not be enough to explain the excess
dileptons observed in this low invariant mass region.

In the future, we need to include properly the expansion dynamics of
the hot matter that is formed in high energy nucleus-nucleus collisions
to obtain a more realistic determination of the dilepton production rate.
We should also include corrections due to the experimental acceptance.
In particular, the transverse momentum cut, applied separately
to the electrons and positrons
in experimental measurements, affects significantly the
dilepton yield near the two pion threshold.
Finally, we have not included the effect of finite baryon density on
the dilepton production rate, which is expected to be
important as well in experiments at the CERN/SPS. We plan to study
consistently both the temperature and the density effect on dilepton
production in high energy heavy ion collisions.

\bigskip

C.S. thanks G. Q. Li for useful conversations about his preliminary results
from the transport model calculation and the baryon density effect on
dilepton production.
The work of C.S. and C.M.K. was supported in part
by the National Science Foundation
under Grant No. PHY-9212209 and the Robert A. Welch Foundation under
Grant No. A-1110.  The work of V.K. was supported by the Director,
Office of Energy Research, Division of Nuclear Physics of the Office of
High Energy and Nuclear Physics of the U.S. Department of Energy under
contract DE-AC03-76SF00098.
The work of S.H.L. was supported by the Basic Science Research
Institute Program, and by  Korea Ministry of Education, BRSI-94-2425,
KOSEF through CTP in Seoul National University.

\newpage
\centerline{{\bf Figure Captions}}

\vspace{1.5cm}

\noindent
{\bf Fig.\ 1:} One loop diagrams for the pion self-energy. The dotted and
solid lines denote the pion and $\rho$ meson, respectively.
\vspace{1.0cm}

\noindent
{\bf Fig.\ 2:} The dilepton production rate from pion-pion annihilation
in hot hadronic matter at $T=160$ MeV. The dotted line is the
result obtained without medium effects while the dashed line is
that including modifications of the pion dispersion relation.
The dash-dotted line is the
production rate obtained by only taking into
account the effect of pion electromagnetic form factor
at finite temperature, and the solid line is the result when
both effects are included.
\vspace{1.0cm}

\noindent
{\bf Fig.\ 3:} The dilepton production rate after integrating over the
dilepton three momentum. The result obtained without medium effects
(dotted line) is compared with
that obtained with modifications of the pion dispersion relation and
electromagnetic form factor at finite temperature (solid line).


\begin{thebibliography}{20}


\bibitem{dilepton} E. L. Feinberg, Nuovo Cimento 34A (1976) 39.

\bibitem{shuryak}  E. V. Shuryak, Phys. Lett. B 78 (1978) 150.

\bibitem{ruu} P. V. Ruuskanen, Nucl. Phys. A 544 (1992) 169c;\\
              J. Kapusta, Nucl. Phys. A 566 (1994) 45c.

\bibitem{kox} C. M. Ko and L. H. Xia, Phys. Rev. Lett. 62 (1989) 1595;\\
              M. Asakawa, C. M. Ko and P. L\'evai,
              Phys. Rev. Lett. 70 (1993) 398;\\
              M. Asakawa and C. M. Ko, Phys. Lett. B 322 (1994) 33.

\bibitem{hadron} K. Kajantie, J. Kapusta, L. McLerran and A. Mekijan,
                 Phys. Rev. D34 (1986) 2746.\\
                 C. Gale and P. Lichard, Phys. Rev D 49 (1994) 3338.\\
                 C. Song , C. M. Ko and C. Gale, Phys. Rev. D 50 (1994) R1827.

\bibitem{dilep-e} G. Agakichiev et. al., CERES Collaboration,
                  CERN preprint CERN PPE/95-26 (1995);\\
                  J. P. Wurm for the CERES Collaboration,
                  in {\it Proc. Quark Matter '95}, January 9-13, 1995;
                  Nucl. Phys. A, to be published.

\bibitem{helios} M. Masera for the HELIOS-3 Collaboration,
                 in {\it Proc. Quark Matter '95}, January 9-13, 1995;
                 Nucl. Phys. A, to be published.

\bibitem{gqli} G. Li, C. M. Ko and G. E. Brown,
               Texas A\&M University Preprint (1995).

\bibitem{denish} D. K. Srivastava, B. Sinha and C. Gale,
                 McGill University Preprint (1995).

\bibitem{MYMA} U.-G. Mei{\ss}ner, Phys. Rep. 161 (1988) 213.

\bibitem{BKY} M. Bando, T. Kugo and K. Yamawaki,
              Phys. Rep. 164 (1988) 217.

\bibitem{slee} S. H. Lee, C. Song and H. Yabu,
               Phys. Lett. B 341 (1995) 407.

\bibitem{song_pion} Chungsik Song, Phys. Rev. D 49 (1994) 1556;
                    Phys. Lett. B 329 (1994) 312.

\bibitem{gale}  C. Gale and J. Kapusta, Phys. Rev. C 35 (1987) 2107.\\
                L. H. Xia, C. M. Ko, L. Xiong, and J. Q. Wu,
                Nucl. Phys. A 485 (1988) 721.

\bibitem{gounaris} G. J. Gounaris and J. J. Sakurai,
                   Phys. Rev. Lett. 21 (1968) 244.

\bibitem{pratt} C. L. Korpa and S. Pratt,
                Phys. Rev. Lett. 64 (1990) 1502.\\
                M. Asakawa, C. M. Ko, P. L\'evai and X. J. Qiu,
                Rev. C 46 (1992) R1159.\\
                M. Herrmann, B. L. Friman and W. N\"orenberg,
                Nucl. Phys, A 560 (1993) 411.

\bibitem{song2} C. Song, S. H. Lee, and C. M. Ko,
                Phys. Rev. C, to be published.

\end{thebibliography}
\end{document}